# A Roof over your Head; House Price Peaks in the UK and Ireland

Peter Richmond
School of Physics
Trinity College
Dublin 2
Ireland

**Abstract**
We analyse, following recent work of Roehner, changes in house prices for both the UK and Ireland. We conclude that prices in the UK/ London have reached a tipping point and house prices relative are set to fall over the next few years. If inflation does not rise then a hard landing seems likely. House prices in Ireland are shown to have broken away from the moderate rise still to found in Northern Ireland and Dublin has emerged as another global 'hot' spot. An evolution similar to that in London can be anticipated

## Introduction

House prices are of concern to many people and at present they seem to be going through one of the periodic up turns. The Financial Times of May 7, 2006 carried on the front page the headline: 'Property Bonanza in Central London' and noted that 'a mini-boom is sweeping the most exclusive streets of London where house prices have risen by 14.5% in a year.' On a different page of the same newspaper there was a rather more cautionary announcement: repossession orders were now at their highest level since the previous peak that occurred as property prices were falling in the early 1990's.

In Dublin house prices that have risen albeit slowly for many decades have been rising very quickly since 1997. One commentator recently reported that 'Frenzy' in the Irish property market has intensified. In the last six months, house price inflation has accelerated to an annual rate of 15%. In Dublin, prices are rising at annual rate of 20%, up from only 3% less than a year ago. But rents have only recently recovered after three-year period in which they were in decline. As a result, yields have been driven down to unprecedented depths.[1]'

Roehner [2] has made an empirical study of house price peaks looking at bubbles that have occurred in the past. Using a simple approach based on work by Sornette[3], he has noted using examples from different times and different parts of the globe that speculative bubbles exhibit certain regularities in shape. Very recently in a new publication he has made a new study of house prices in the Western USA and made a prediction as to the way these house prices will evolve over the next few years [4].

In this note we develop the idea and apply it to house prices in the UK over the period 1952-2006. As a result we conclude that a similar speculative house price bubble is currently under way in the UK. The data allows a similar interpretation to that of Roehner for the US and as to the evolution of UK house prices over the next few years is possible. In addition we examine data for Irish house prices over the period 1997-2006. By comparing this data with that for the UK we can see that Dublin appears to have joined the 'global' group of cities that exhibit similar house price dynamics and that for the first time in recent years, house price dynamics are synchronized with London. We can anticipate a similar evolution in the future.

## House prices in the UK

House price data for the UK and Northern Ireland can be obtained from both the Halifax and Nationwide Building society web sites. Both are similar and figure 1 below has been downloaded from the Nationwide Building Society.

Fig 1 House prices (£) in the United Kingdom (1973-2006)

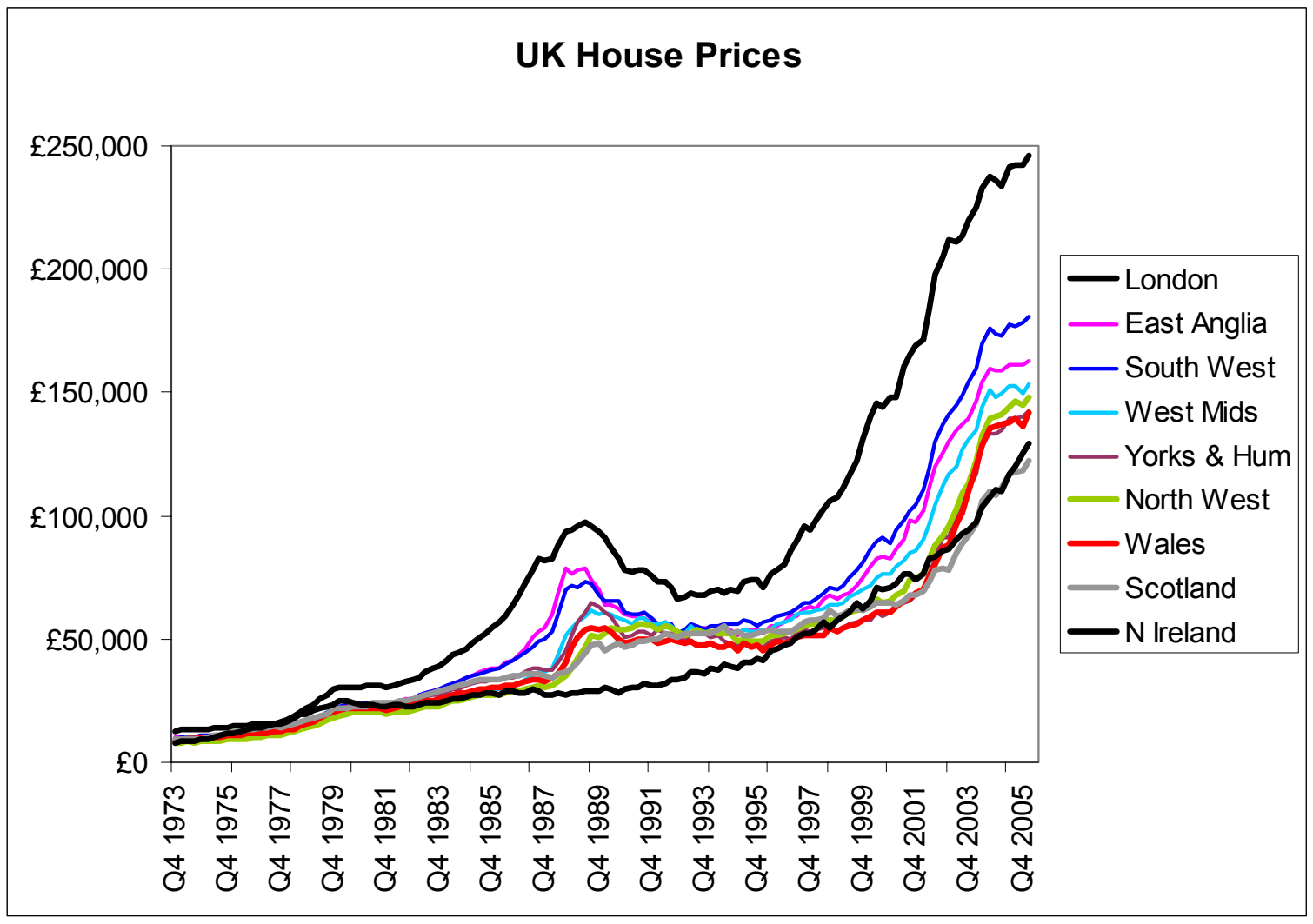

The peak that occurred in the UK during the late 1980's is clearly visible. The geographical variation across the UK has been studied by Roehner. We note here that lower line for Northern Ireland that is essentially Belfast. This is apparently uncorrelated with the frenetic activity on the UK mainland and we return to this point later.

However following Roehner it is more revealing to compute the house prices in constant money terms. This is shown in figure 2.

Fig 2: Log (house price/RPI ) for London over the period 1952-2006

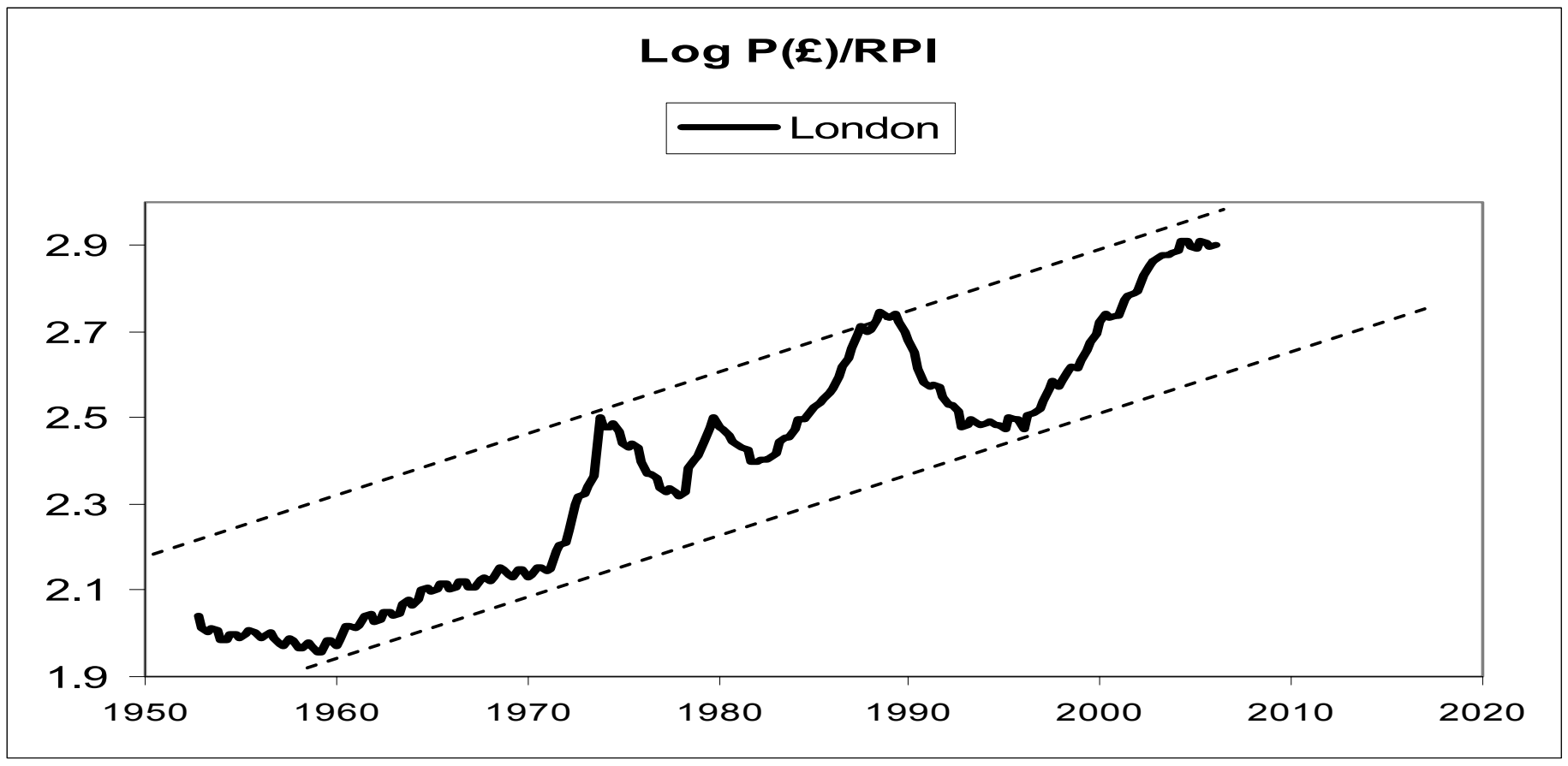

It is interesting to compare this graph with that recently published by Roehner for house prices in the Western US. This is shown below in Fig 3.

Fig 3 Log house prices ($US) in the Western US relative to the retails price index computed by Roehner [4]

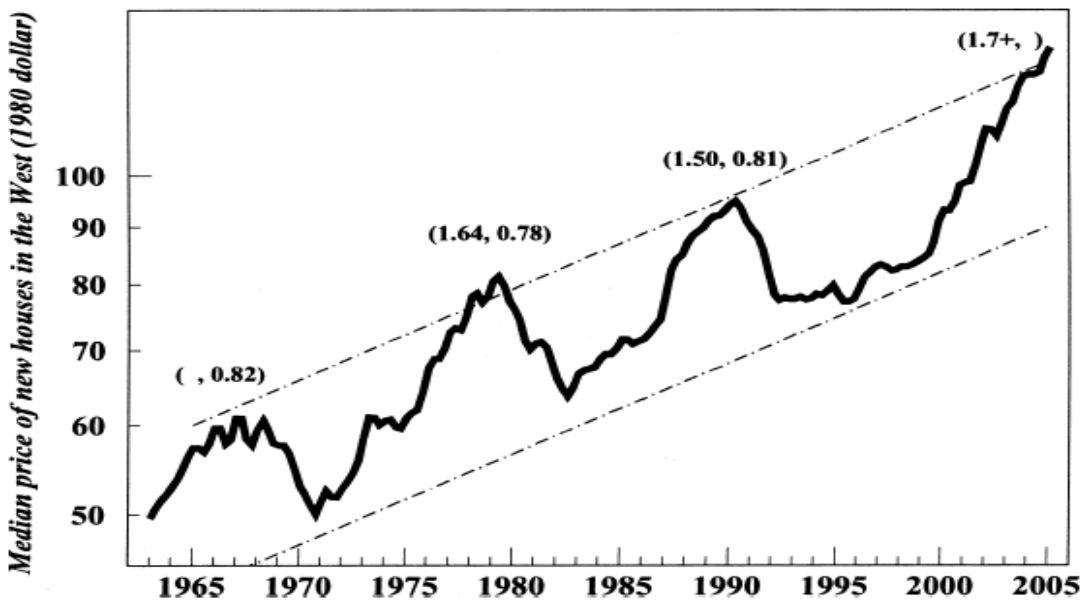

The US data shows strong peaks at ~1969, 1979 and 1989. Clear peaks on the UK data are visible at 1974, 1979 and 1989. Looking closely at the US data we see some evidence of a peak or shoulder at ~1973. This was ignored by Roehner, who focussed on the almost identical character of the major peaks, but now we can see that it corresponds to a similar but more pronounced peak that occurred in London at this time. We note here that Roehner used smoothed data points that possibly reduced the impact of some of these secondary peaks. We did not smooth the data used to compute the curve in Fig 2. In passing, this author recalls that Australia experienced a similar boom during the period 1973-4 from which he profited as he relocated from Australia back to the UK. Looking even more closely at the UK data there is a suggestion of a very weak peak around 1969 but without the supporting evidence of the strong peak in the US at this time, one would probably ignore this. However, all this suggests house price correlations have increased not only between the UK and the US since World War 2 but also across the English speaking world. At this time there is essentially complete correlation between the UK and the US. Furthermore from the UK data it seems clear that the house price peak is not only at a maximum but is tipping over. This is consistent not only with the work of Roehner but also Sornette and Zhow [5] who have also predicted using more similar albeit complex arguments based on the existence of log periodic oscillations that the US data will peak in 2006.

In figure 4 we compare the US Dow Jones with the FTA index since the late 19th century and one can see that the US economic engine began to develop strongly after the Civil war. The UK index however did not really get into gear until after the Second World War. By the 1980's it is clear that there is essentially complete correlation between the two indices.

Figure 4 Graph of the FTA (1800-1990) and Dow Jones (1890-1990)

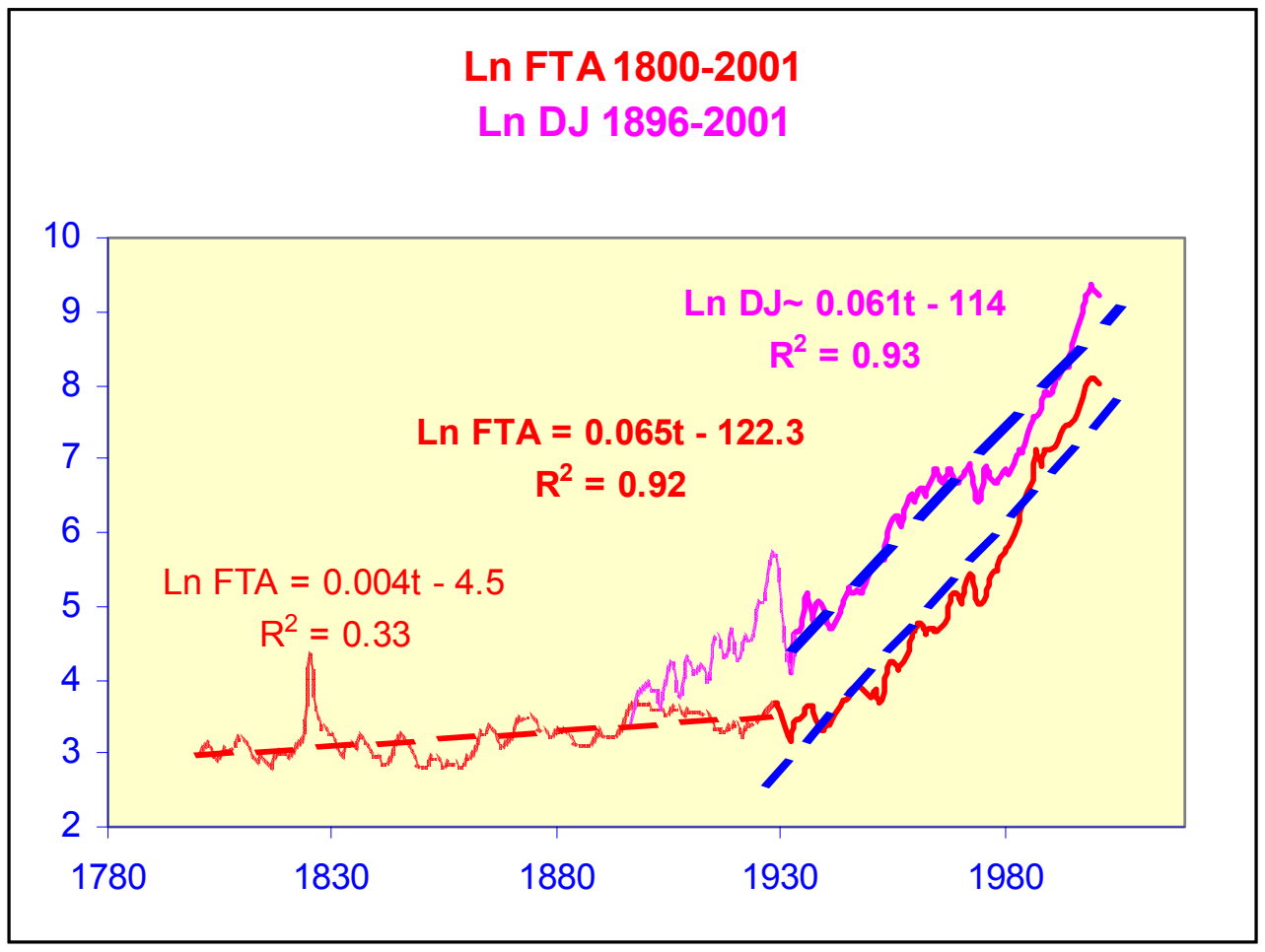

If one assumes that the mechanism driving house prices continues now in a manner similar to activity since the 1980's in the UK, or the 1960's in the US, then one can assert on the basis of the above evidence that the peak has been reached and is now tipping over into the downturn. If it follows past trends then we can expect a hard landing with falls of 6-8% per annum relative to inflation for then next 4-5 years. Whether it will continue and the full triangular shape of the current peak takes form we cannot say with certainty. Perhaps governments will have some hitherto unknown policies or tools that they may implement to counteract such a downturn but on again on the basis of past experience it seems unlikely, although a strong dose of inflation might help! The author was interested to see that at the time of writing this piece, the BBC News web site was carrying the headline '**Markets beset by inflation fears'[6].** That being said perhaps by the time of the next general election in the UK, the government might be able to say that housing is once again available for all at reasonable prices!

**House prices in Ireland**

Long time series for house price data comparable to those available for the UK seem not to be readily available. Figure 5 shows

Fig 5 House prices in the Republic of Ireland (1997-2006)

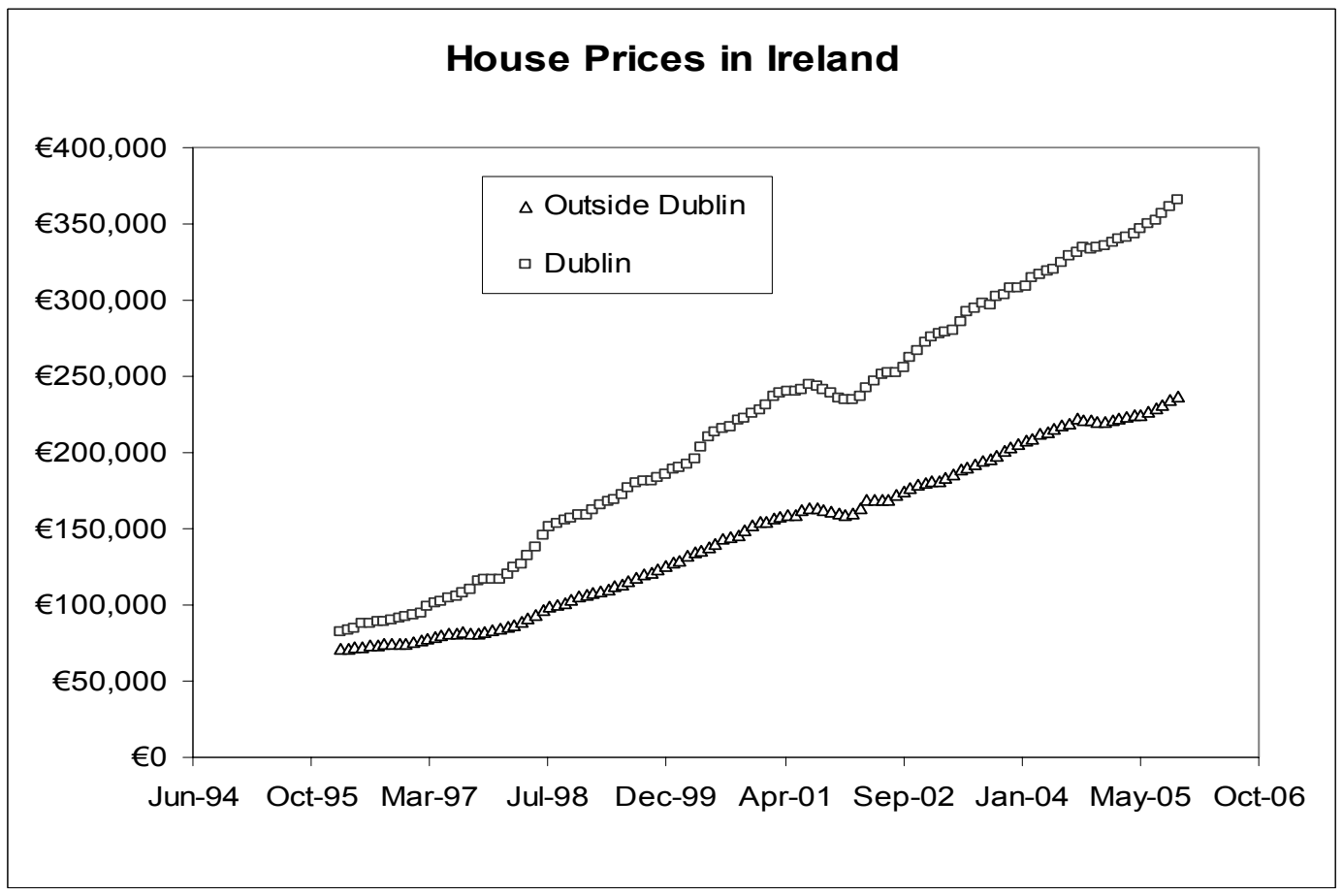

data for prices of houses both in Dublin and outside for the period 1995 to the first quarter of 2006. Both curves are rising strongly over this period in line with both US and UK data. Can we say any more? As a first step we have converted the data from EUR into GBP and superimposed the data onto that for the UK and Northern Ireland (Figure 5)

Fig 6 House prices in the Republic of Ireland converted to £ and superposed on similar graphs for London and Northern Ireland.

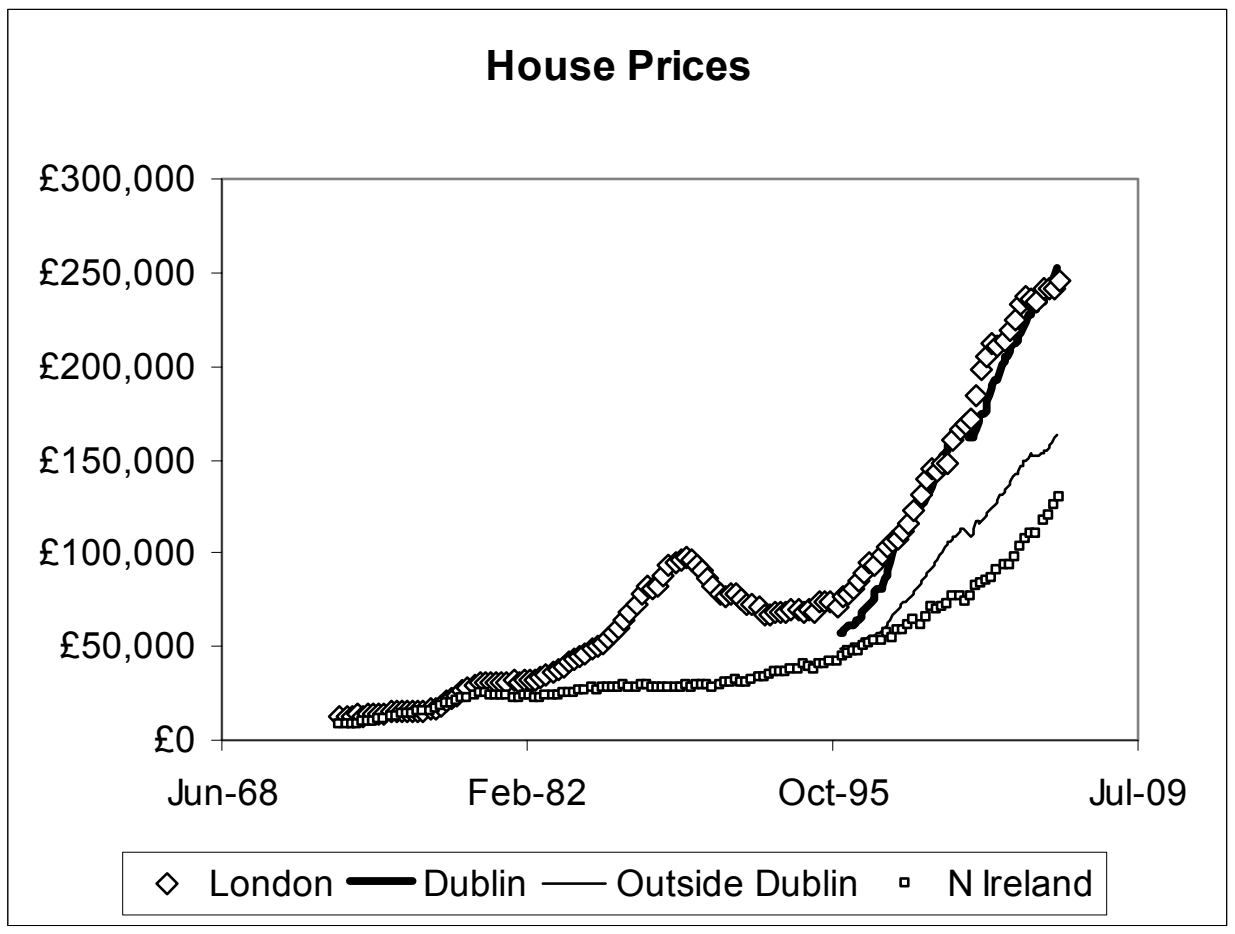

It is immediately evident that the data over this period has broken away from Northern Ireland. Data for Dublin is now essentially following that of London and the data for houses outside Dublin is following that of a provincial city in England.

We can go further and, convert the Irish data using the Irish consumer price index to constant 1980 prices. The superposition of this data for Dublin onto similar prices for London is shown in figure 7.

Fig 7 Dublin house prices (Log £/Irish consumer price index) relative to the similar ration for London.

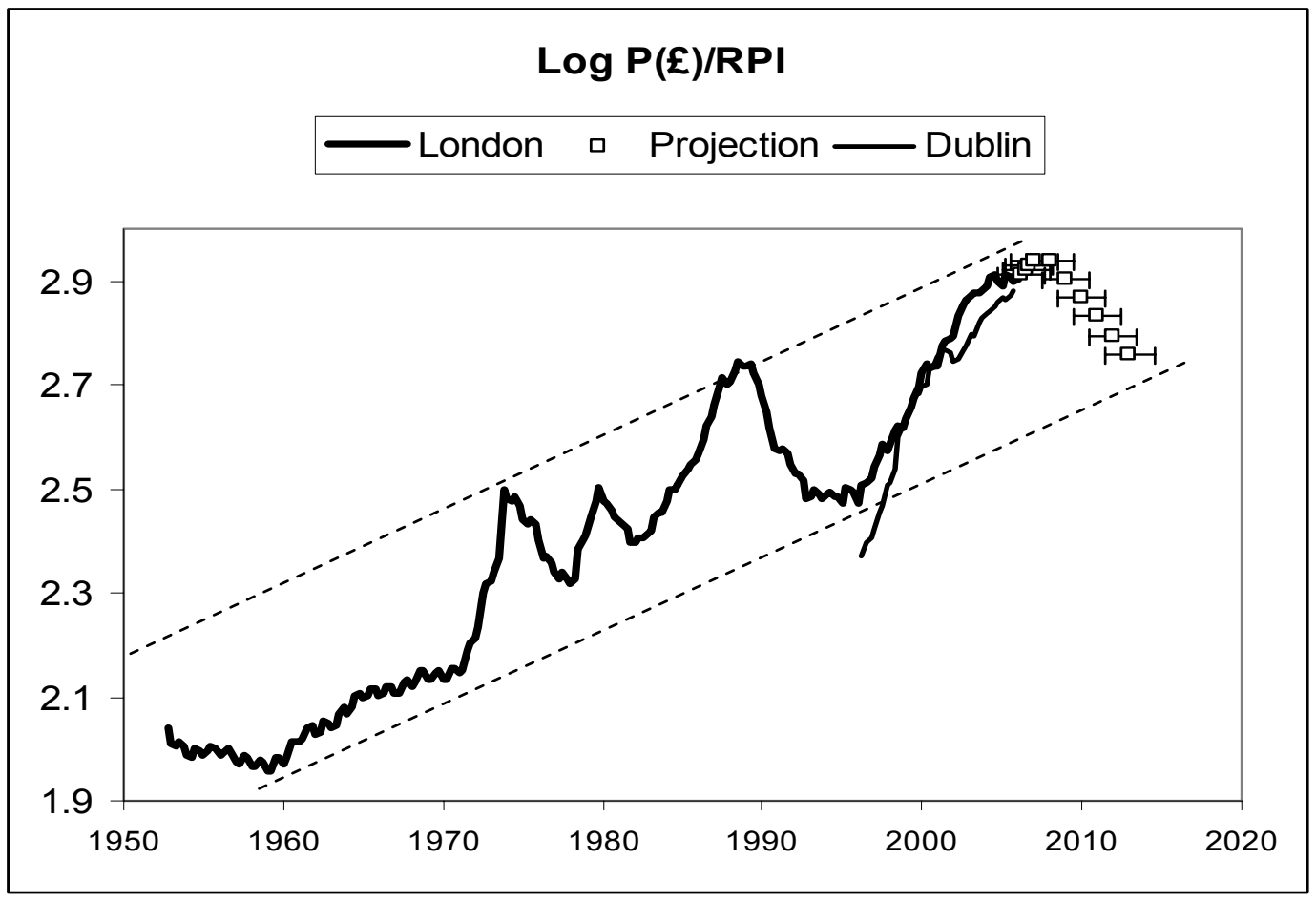

It seems that the Irish prices have also peaked. Perhaps the downturn will lag a little behind that for the UK according to the projection we have made for prices in London now shown in open squares. However a downturn there will be according to these calculations.

It seems clear that Dublin as a result of the economic expansion during the last 20 years has now emerged as a major city on a global scale and is also driving house prices in the manner of all other major cities.

**Conclusions**

House prices, it would seem, exhibit regular, almost universal behaviour across time and geographic domain. Roehner has suggested that major cities are 'hot spots' where this activity is galvanised. From the perspective of the physicist, this suggests that mechanisms akin to self organised criticality are driving such asset prices.

As a result of substantial investment and economic activity in recent decades, Dublin now seems to have joined this group of 'hot spots'. The consequence is that prices will become rather more volatile than in the past exhibiting peaks and troughs that are the signature of other 'hot spots'. This increase in correlation of house prices between cities is very reminiscent of phase locking in complex systems observed recently by Holyst [7]

Whilst house prices in all these cities across the world have been rising strongly in recent years, the analysis suggests that they have now peaked and an equally strong downturn (relative to inflation) is likely.

**Acknowledgement**
I am grateful to Bertrand Roehner of the Institute for Theoretical and High Energy Physics, University Paris 7 for many discussions, comments and helpful suggestions as this work was being developed.

---

[1] Rossa White Davy Research Dublin 29th March 2006 Research note.
[2] B M Roehner Hidden Collective Factors in Speculative Trading, Springer, Berlin, 2001, ISBN 3 540 41294 8
[3] A Johansen, D Sornette and O Ledoit, Predicting Financial Crashes using Discrete Scale Invarience, Journal of Risk, 5 (1999) 5-32 and cond-mat/9903321
[4] B M Roehner Real Estate Price Peaks - A Comparative Overview. Evol. Inst. Econ. Rev. 2(2): 167–182 (2006)
[5] W-X Zhou and D Sornette, Is There a Real Estate Bubble in the US? arXiv: physics/0506027 v1 3 June 2005
[6] Markets beset by inflation fears http://news.bbc.co.uk/1/hi/business/5054484.stm
[7] D. Helbing, M. Schonhof, H.U. Stark and J.A. Holyst, How individuals learn to take turns: Emergence of alternating cooperation in a congestion game and the prisoner's dilemma, Adv. Complex Systems 8, 87 (2005)